\documentclass[a4paper]{article}
\usepackage{amsmath,amsfonts,amssymb,amsthm}

\date{November 3, 1998}
\newcommand{\field}[1]{\mathbb{#1}}
\newcommand{\N}{\field{N}}
\newcommand{\Z}{\field{Z}}
\newcommand{\R}{\field{R}}
\newcommand{\C}{\field{C}}
\newcommand{\D}{\mathfrak{D}}

\newcommand{\E}{\mathcal{E}}
\newcommand{\F}{\mathcal{F}}

\newcommand{\vektor}[1]{\boldsymbol{#1}}
\newcommand{\ve}{\vektor{e}}
\newcommand{\vx}{\vektor{x}}
\newcommand{\vy}{\vektor{y}}
\newcommand{\vk}{\vektor{k}}
\newcommand{\vp}{\vektor{p}}
\newcommand{\vA}{\vektor{A}}
\newcommand{\vJ}{\vektor{J}}
\newcommand{\vR}{\vektor{R}}
\newcommand{\vn}{\vektor{n}}
\newcommand{\valpha}{\vektor{\alpha}}
\newcommand{\vomega}{\vektor{\omega}}
\newcommand{\vsigma}{\vektor{\sigma}}

\renewcommand{\H}{\mathcal{H}}

\newcommand{\eps}{\varepsilon}

\newcommand{\const}{\mbox{const}}
\newcommand{\supp}{\mbox{supp}}

\newcommand{\sprod}[2]{\langle #1,#2 \rangle}       

\newtheorem{theorem}{Theorem}
\newtheorem{lemma}[theorem]{Lemma}
\newtheorem{corollary}[theorem]{Corollary}

\begin{document}

\title{Instability of a Pseudo-Relativistic Model of Matter with Self-Generated Magnetic Field}

\author{Marcel Griesemer\footnote{Present address:
     Department of Mathematics, University of Alabama at Birmingham,
    Birmingham, AL 35294, USA.} and Christian Tix\\Mathematik,
  Universit\"at Regensburg\\D-93040 Regensburg}
\date{}

\maketitle
\begin{abstract}
  For a pseudo-relativistic model of matter, based on the
  no-pair Hamiltonian, we prove that the inclusion
  of the interaction with the self-generated magnetic
  field leads to instability for all positive values of the fine
  structure constant.  This is true no matter whether this interaction is
  accounted for by the Breit potential, by an external magnetic field
  which is chosen to minimize the energy, or by the quantized
  radiation field.
\end{abstract}
\bigskip


\section{Introduction}\label{sec1}

The stability of matter problem concerns the question whether the
minimal energy of a system of particles is bounded from below
(stability of the first kind), and whether it is bounded from
below by a constant times the number of particles (stability of
the second kind). Stability of the second kind for
non-relativistic quantum-mechanical electrons and nuclei was first
proved in 1967 by Dyson and Lenard \cite{DysonLenard1967I,
DysonLenard1967II}. Since the new proofs of Lieb and Thirring, and
Federbush in 1975 stability of matter is a subject of ongoing
interest dealing with more and more realistic models of matter
such as systems with a classical or quantized magnetic field
included or with relativistic electrons (see \cite{Liebetal1997}
and the references therein). Stability with relativistic electrons
is more subtle because of the uniform 1/length scaling behavior of
the energy, which holds for massless particles (high
particle-energy limit). Then the minimal energy is either
non-negative or equal $-\infty$, so that stability of the second
kind becomes equivalent to the statement that stability of the
first kind holds for any given number of particles. We simply call
this stability henceforth.

This paper is about a pseudo-relativistic model of matter which is
stable, but which becomes unstable when the electrons are allowed
to interact with the self-generated magnetic field.  The
self-generated magnetic field may be described using either an
effective potential (the Breit-potential), an external magnetic
field over which the energy is minimized, or the quantized
radiation field. In all these cases we find instability for all
positive values of the fine-structure constant. In contrast to
most other models, where the collapse of the system, if it occurs,
is due to the attraction of electrons and nuclei
\cite{Liebetal1986, LiebLoss1986, LiebYau1988, Loss1997} (there
would be no collapse without this interaction), the instability
here is due to the attraction of parallel currents.

The model we study is based on a pseudo-relativistic Hamiltonian
sometimes  called no-pair or Brown-Ravenhall Hamiltonian
describing $N$ relativistic electrons and $K$ fixed nuclei
interacting via Coulomb potentials. The electrons are vectors in
the positive energy subspace of the free Dirac operator and their
kinetic energy is described by this operator. For a physical
justification of this model see the papers of Sucher
\cite{Sucher1980, Sucher1987}, for applications of the model in
computational atomic physics and quantum chemistry see Ishikawa
and Koc \cite{IshikawaKoc1994, IshikawaKoc1997}, and
Pyykk\"o~\cite{Pyykkoe1988}. The Brown-Ravenhall Hamiltonian
yields stability for sufficiently small values of the fine
structure constant and the charge of the
nuclei~\cite{Evansetal1996,Tix1997, Tix1997b,
BalinskyEvans1998b,Liebetal1997}; there are further rigorous
results concerning the virial theorem~\cite{BalinskyEvans1998a}
and eigenvalue estimates~\cite{GriesemerSiedentop1997}.

We are interested in the minimal energy of this model when it is
corrected to account for the interaction of the electrons with the
self-generated magnetic field. This correction may be done for
instance by introducing a external magnetic field
$\nabla\times\vA$ to which the electrons are then minimally
coupled and whose field energy is added to the energy of the
system. The field $\vA$ is now considered part of the system and
hence the energy is to be minimized w.r.t. $\vA$ as well. The
minimizing $\vA$ for a given electronic state is the
self-generated one (to avoid instability for trivial reasons the
gauge of $\vA$ has to be fixed). The energy of this system is
unbounded from below if $N\alpha^{3/2}$ is large, $\alpha$ being
the fine structure constant, even if the vector potential is
restricted to lie in a two parameter class
\(\{\gamma\vA_0(\delta\vx):\gamma,\delta\in \R_+\}\) where $\vA_0$
is fixed and obeys a weak condition requiring not much more than
$\vA_0\not\equiv 0$. This is our first main result.  It extends a
previous result of Lieb et al. \cite{Liebetal1997} and is
reminiscent of the fact that a static non-vanishing classical
magnetic field in QED is not regular, in the sense that the
dressed electron-positron emission and absorption operators do not
realize a representation of the CAR on the Fock space of the free
field \cite{NenciuScharf1978}.

Alternatively the energy-shift due to the self-generated magnetic
field may approximately  be taken into account by including the
Breit potential in the energy. The resulting model is unstable as
well. That is, the energy is unbounded from below if
$N\alpha^{3/2}$ is large, no matter how small $\alpha$ is. This is
our second main result. It concerns a Hamiltonian that is closely
related the Dirac-Coulomb-Breit or Dirac-Breit Hamiltonian, which
is the bases for most calculations of relativistic effects in many
electron atoms \cite{Sucher1980, Pyykkoe1988}. We mention that for
$\alpha=1/137$ the energy is bounded below if $N\leq 39$ and
unbounded below if $N\geq 3.4\cdot 10^7$ (Theorem~\ref{theorem3}
and Theorem~\ref{theorem2}).

A third way of accounting for the self-generated field is to
couple the electrons to the quantized radiation field. From a
simple argument using coherent states (Lemma~\ref{lemma}) it
follows that the instability of this model is rather worse than
the instability of the model first discussed.

As mentioned above the instability with external magnetic field
was previously found by Lieb et al. \cite{Liebetal1997}. Our
result extends their result and our proof is simpler. The model
with Breit interaction corresponds to the classical system
described by the Darwin Hamiltonian, which has been studied in the
plasma physics literature (see \cite{AppelAlastuey1998} an the
references therein). This classical model is thermodynamically
unstable as well \cite{Kiessling1998}.

In Sections~\ref{sec2}, \ref{sec3} and \ref{sec4} we introduce the
models with external magnetic field, with Breit potential, and
with quantized radiation field and prove their instability
(Theorem~\ref{theorem1}, Theorem~\ref{theorem2}, and
Lemma~\ref{lemma}). In Section~\ref{sec3} we also discuss dynamic
nuclei for the model with Breit potential. There is an appendix
where numerical values for stability bounds on $N\alpha^{3/2}$
given in the main text are computed.

\section{Instability with Classical Magnetic Field}\label{sec2}

We begin with the model of matter with external magnetic field. For
simplicity the electrons are assumed to be non-interacting and no nuclei
are present. We could just as well treat a system of interacting
electrons and static nuclei and would obtain essentially
the same result (see Remark 4 below).

Consider a system of $N$ non-interacting electrons in the external
magnetic field \(\nabla\times \vektor{A}\). The energy of this system
is
\begin{equation*}
\E_N(\psi,\vA)=\sprod{\psi}{\sum_{\mu=1}^{N}D_{\mu}(\vektor
    {A})\psi}+\frac{1}{8\pi}\int |\nabla\times\vA(\vx)|^2d\vx
\end{equation*}
where \(D_{\mu}(\vA)\) is the Dirac operator
\(D(\vA)=\valpha\cdot(-i\nabla +\alpha^{1/2}\vA(\vx))
+ \beta m\) acting on the $\mu$-th particle, and the vector $\psi$,
describing the state of the electrons, belongs to the Hilbert space
\begin{eqnarray*}
  \H_N &=& \bigwedge_{\mu=1}^N \Lambda_{+}L^2(\R^3,\C^4)\\
  \Lambda_{+} &=& \chi_{(0,\infty)}(D(\vA\equiv 0)),
\end{eqnarray*}
or rather the dense subspace \(\D_N=\H_N\cap H^1[(\R^3\times\{1,\ldots,4\})^N]\).
That is, an electron is by definition a vector in the positive energy
subspace of the free Dirac operator. We will always assume the
vector potential $\vA$ belongs to the class $\mathcal{A}$ defined by
the properties:
\begin{eqnarray*}
i) & & \nabla\cdot\vA = 0,\\
ii)& & \vA(\vx) \rightarrow 0\makebox[2cm]{as}|\vx|\rightarrow \infty,\\
iii) & & \int_{\R^3}|\nabla\times\vA|^2 < \infty.
\end{eqnarray*}
Notice that $\H_N$ is not invariant under multiplication with
smooth functions, in particular it is not invariant under gauge
transformations of the states. It follows that the minimal energy
for fixed $\vA$ is gauge-dependent. It can actually be driven to
$-\infty$ by a pure gauge transformation (see Remark 3 below). To
avoid this trivial instability we fixed the gauge of $\vA$ by
imposing conditions i) and ii).

The constants $\alpha>0$ and $m\geq0$ in the definition of
\(D(\vA)\) are the fine structure constant and the mass of the
electron respectively. In our units \(\hbar=1=c\), so that
\(\alpha =e^2\) which is about $1/137$ experimentally. We denote the
Fourier transform of a function $f$ by $\widehat{f}$ or $\mathfrak{F}(f)$ and use $\vp$ or
$\vk$ for its argument rather then $\vx$ or $\vy$. Our first result is

\begin{theorem}\label{theorem1}
  Suppose $\vA\in \mathcal{A}$ is such that \(Re[\ve\cdot
  \widehat{\vA}(\vp)] < 0\) in $B(0,\eps)$ for some $\ve\in\R^3$ and
  $\eps>0$. Then there exist a constant $C_{\vA}$ such that for all
  $\alpha > 0$, $m\geq 0$ and \(N\geq C_{\vA}\alpha^{-3/2}\)
\[ \inf_{\psi\in\D_N,\|\psi\|=1,\ \gamma,\delta\in \R_{+}} \E_N(\psi,\gamma\vA(\delta\vx))= -\infty.\]
\end{theorem}

\noindent
{\em Remarks.}
\begin{enumerate}
\item It is sufficient that \(\vA\in \mathcal{A}\cap L^1\) and
  \(\int_{\R^3}\vA(\vx)d\vx \neq 0\), since $\widehat{\vA}$ is then
  continuous and \(\widehat{\vA}(0)\neq 0\). Thus we have instability
  for virtually all non-vanishing $\vA\in\mathcal{A}$.
\item The smallness of $N\alpha^{3/2}$ is not only necessary but also
  sufficient for stability (see \cite[Section 4]{Liebetal1997}).
\item If the condition ii) that $\vA$ vanishes at infinity (and thus
  the gauge fixing) is dropped there is instability even for $N=1$ and
  the theorem becomes trivial. In fact for $N=1$ and \(\vA(\vx)\equiv
  \vektor{a}\neq 0\), \(\E_{N=1}(\psi,\gamma\vA) =
  \sprod{\psi}{D(0)\psi}+ \gamma\alpha^{1/2}\vektor{a}\int
  \psi^+(\vx)\valpha\psi(\vx)d\vx\) which, as a function of $\gamma$,
  is unbounded from below for suitable
  \(\psi\in\Lambda_{+}L^2(\R^3,\C^4)\).
\item The statement of the theorem also holds for the system of
  electrons and static nuclei with energy
  \(\E_N(\psi,\vA)+\alpha\sprod{\psi}{V_c\psi}\) where
  \begin{equation}\label{Coulomb}
    V_c:=
    -\sum_{\mu=1}^{N}\sum_{\kappa=1}^{K} \frac{Z_{\kappa}}{|\vx_{\mu}-
      \vR_{\kappa}|}
    + \sum_{\mu <\nu}^{N}\frac{1}{|\vx_{\mu}- \vx_{\nu}|} +
    \sum_{\kappa<\sigma}^{K} \frac{Z_{\kappa}
      Z_{\sigma}}{|\vR_{\kappa}-\vR_{\sigma}|}
  \end{equation}
  if both $N$ and $\sum Z_{\kappa}$ are bigger than
  \(C_{\vA}\alpha^{-3/2}\) and if the energy is in addition minimized
  with respect to the pairwise distinct nuclear positions
  $\vR_{\kappa}$. (see the proof of Theorem~\ref{theorem2}).
\item Quantizing the radiation field does not improve the stability of
  the system (see Section~\ref{sec4}).
\end{enumerate}

The only way to restore stability we know is to replace $\H_N$
  by the $\vA$-dependent Hilbert space
  \[ \H_{N,\vA}= \bigwedge_{\mu=1}^N \chi_{(0,\infty)}(D(\vA))L^2(\R^3,\C^4).\]
  Obviously \(\E_N(\psi,\vA)\geq 0\) for \(\psi\in \H_{N,\vA}\). In
  fact even \(\E_N(\psi,\vA)+\alpha\sprod{\psi}{V_c\psi}\) is
  non-negative for $Z_{\kappa}$ and $\alpha$ small enough
  \cite{Liebetal1997}.

\begin{proof}[Proof of Theorem~\ref{theorem1}] We will only work with
Slater determinants and the following representation of
one-particle orbitals. If \(u\in L^2(\R^3;\C^2)\) then
\begin{equation}\label{apf1}
  \widehat{\psi}(\vp)=\left(\frac{E(\vp)+m}{2E(\vp)}\right)^{1/2}
  \begin{pmatrix}u(\vp)\\ \frac{\vsigma\cdot\vp}{E(\vp)+m}u(\vp)
  \end{pmatrix},
\end{equation}
with \(E(\vp)=\sqrt{\vp^2+m^2}\), is the Fourier
transform of a vector $\psi\in\Lambda_{+}L^2$, and the map
\(u\mapsto\psi,
L^2(\R^3;\C^2)\rightarrow\Lambda_{+}L^2(\R^3;\C^4)\) is unitary.

It suffices to consider the case $m=0$ and find a Slater determinant
\(\psi=\psi_1\wedge\ldots\wedge\psi_N\) and $\gamma,\delta\in \R_{+}$ such
that \(\E_N(\psi,\gamma\vA(\delta\vx))<0\). In fact by the scaling
\(\psi\mapsto\psi_{\delta},\ \vA\mapsto\vA_{\delta}\) defined by
\(u_{\mu,\delta} =\delta^{-3/2}u_{\mu}(\delta^{-1}\vp)\) and
\(\vA_{\delta}(\vx) =\delta\vA(\delta\vx)\) we can then drive the
energy with $m>0$ to $-\infty$ because \(\E(\psi_{\delta},\vA_{\delta},m)=
\delta\E(\psi,\vA,m/\delta)\) and \(\E(\psi,\vA,m/\delta)\rightarrow
\E(\psi,\vA,m=0)\) for \(\delta\rightarrow\infty\).

{\em Choice of $\psi$.} Let $Q$ be the unit cube \(\{\vp\in\R^3|0\leq
p_i\leq 1\}\), \(u(\vp)=(\chi_Q(\vp), 0)^T\), and $\ve\in\R^3$ an
arbitrary unit vector. Set
\begin{equation}\label{apf1.5}
 u_{\mu}(\vp)= u(\vp-\lambda N^{1/3}\ve-\vn_{\mu}),\hspace{3em}
\mu=1,\ldots,N
\end{equation}
where $\lambda$ is a positive constant to be chosen sufficiently large
later on, and \((\vn_{\mu})_{\mu=1\ldots N}\subset\Z^3\) are
the $N$ lattice sites nearest to the origin, i.e.,
\(\max_{\mu=1\ldots N}|\vn_{\mu}|\) is minimal. We define
\(\psi=\psi_1\wedge\ldots\wedge\psi_N\) by
\begin{equation}\label{apf2}
 \widehat{\psi}_{\mu}(\vp)=\frac{1}{\sqrt{2}}\begin{pmatrix}u_{\mu}(\vp)\\
 \vsigma\cdot\vomega_{\vp}u_{\mu}(\vp)
 \end{pmatrix},\hspace{2em} \vomega_{\vp}=\frac{\vp}{|\vp|},
\end{equation}
which is (\ref{apf1}) for $m=0$. Then $\psi\in\H_N$ and
\(\sprod{\psi_{\mu}}{\psi_\nu}= \sprod{u_{\mu}}{u_\nu} =
\delta_{\mu\nu}\). Notice that
\begin{equation}\label{apf3}
  |\vp-\lambda N^{1/3}\ve|\leq N^{1/3}\makebox[5em]{for all}\vp\in\supp(u_{\mu})
\end{equation}
at least for large $N$ (see the appendix), i.e., in Fourier space all
electrons are localized in a ball with radius $N^{1/3}$ and a distance
from the origin which is large compared to the radius (since $\lambda$
will be large).

Since \(\psi=\psi_1\wedge\ldots\wedge\psi_N\) and $m>0$ we have
\begin{equation}\label{apf4}\begin{split}
  \E_N(\psi,\vA) = &\sum_{\mu=1}^{N}\sprod{\psi_{\mu}}{|\nabla|\psi_{\mu}} +
  \alpha^{1/2}\sum_{\mu=1}^{N}\int \vJ_{\mu}(\vx)\vA(\vx)d\vx\\
  & + \frac{1}{8\pi}\int |\nabla\times\vA(\vx)|^2d\vx
\end{split}\end{equation}
where \(\vJ_{\mu}(\vx)= \psi_{\mu}^{*}(x)\valpha \psi_{\mu}(x)\).
By definition of \(\psi_{\mu}\)
\begin{equation}\label{apf5}
  \widehat{\vJ}_{\mu}(\vp)=
  \frac{1}{2}(2\pi)^{-3/2}\int
  u_{\mu}^{\ast} (\vk-\vp)
  \left[\vsigma
  (\vomega_{\vk}\cdot\vsigma) +
  (\vomega_{\vektor{k-p}}\cdot\vsigma)\vsigma
  \right]u_{\mu}(\vk)d\vk.
\end{equation}
Replace here $u_{\mu}$ by its defining expression and substitute
\((\vk-\lambda N^{1/3}\ve -\vn_{\mu}) \mapsto\vk\). Since
\(\vomega_{\vk+\lambda N^{1/3}\ve+\vn_{\mu}} \rightarrow\ve\) as
\(\lambda\rightarrow\infty\) and since $u$ has compact support, it
follows that \(\widehat{\vJ}_{\mu}(\vp)\) converges to the current
\begin{equation}\label{apf6}
  \widehat{\vJ}_{0}(\vp)= \ve\,(2\pi)^{-2/3}\int
  u^{\ast} (\vk-\vp) u(\vk)d\vk
\end{equation}
as $\lambda\rightarrow\infty$. More precisely
\(|\widehat{\vJ}_{\mu}(\vp)-\widehat{\vJ}_{0}(\vp)|\leq
C\lambda^{-1} |\widehat{\vJ}_{0}(\vp)|\) for \(\lambda\geq\lambda_0\)
where $\lambda_0$ and $C$ are independent of $\mu$ and $N$. From
\(\widehat{\vJ}_{0}(\vp)|\vp|^{-1},\ \widehat{\vA}(\vp)|\vp|\in L^2\)
it follows that
\begin{equation}\label{apf7}
\int\widehat{\vJ}_{\mu}^{\ast}(\vp)\widehat{\vA}(\vp)d\vp  =
\int\widehat{\vJ}_{0}^{\ast}(\vp)\widehat{\vA}(\vp)d\vp +
O(\lambda^{-1}), \hspace{2em}\lambda\rightarrow\infty.
\end{equation}
After a scaling \(\vA\mapsto\vA_{\delta}\) we may assume \({\mathrm
  Re}[\ve\cdot\widehat{\vA}(\vp)]<0\) in the support of
$\widehat{\vJ}_{0}$ rather then in \(B(0,\eps)\), so that
(\ref{apf7}) is bounded from above by some $-c_1<0$ for
$\lambda\geq\lambda_0$ where $c_1$ and $\lambda_0$ are independent of
$\mu$ and $N$. Observing finally that
\begin{equation}\label{apf8}
 \sprod{\psi_{\mu}}{|\nabla|\psi_{\mu}} = \int |\widehat{\psi}_{\mu}(\vp)|^2|\vp|d\vp
 \leq (\lambda+1)N^{1/3}
\end{equation}
for all $\mu$, we conclude
\begin{eqnarray*}
  \E_N(\psi,\gamma\vA) &\leq&  (\lambda_0+1)N^{4/3} - \alpha^{1/2}\gamma N
  c_1 + \gamma^2 c_2\\
  &=& (\lambda_0+1)N^{4/3} - \alpha \frac{c_1^2}{4c_2}N^2
\end{eqnarray*}
which is negative for $N\alpha^{3/2}$ large enough. At the end we
inserted the optimal $\gamma$.
\end{proof}

The theorem has the obvious corollary

\begin{corollary}\label{corollary}
  There is a constant $C$ such that for all $\alpha>0$, $m\geq 0$ and \(N\geq C\alpha^{-3/2}\),
  \[ \inf_{\psi\in\D_N,\|\psi\|=1; \vA\in\mathcal{A}} \E_N(\psi,\vA) = -\infty.\]
\end{corollary}

This result is due to Lieb, Siedentop, and Solovej
\cite{Liebetal1997}.

\noindent {\em Remark.} It is sufficient that \(C=1.4\cdot 10^5\)
or that \(N\geq 3.4\cdot 10^7\) for \(\alpha^{-1}=137\), see the
appendix.\\

To conclude this section we compute
\(\min_{\vA\in\mathcal{A}}\E_N(\psi,\vA)\). This will provide a link
to the instability with Breit-potential discussed in the next section. To
exhibit the $\vA$-dependence we write the energy as
\begin{equation*}
  \E_N(\psi,\vA) = \E_N(\psi,\vA\equiv 0) + \alpha^{1/2} \int
  \vJ(\vx)\vA(\vx) + \frac{1}{8\pi}\int |\nabla\times\vA(\vx)|^2d\vx,
\end{equation*}
where $\vJ(\vx)$ is the probability current density associated
with $\psi$. Its functional dependence on $\psi$ is not crucial
here. A straight forward computation shows that the Euler-Lagrange
equation for $\vA$ is \(-\Delta \vA = 4\pi \alpha^{1/2} \vJ_T\)
where $\vJ_T$ is the divergence free - or transversal - part of
$\vJ$. Comparing this equation with the Maxwell-equation for $\vA$
in Coulomb gauge, which is \(\square \vA = 4\pi\alpha^{1/2}
\vJ_T\), we find that the minimizing magnetic field is the
self-generated one up to effects of retardation. Solving the
Euler-Lagrange equation gives
\begin{equation}\label{minimum}
  \min_{\vA\in\mathcal{A}}\E_N(\psi,\vA) = \E_N(\psi,\vA\equiv 0) - \frac{\alpha}{2}\int
  \frac{\vJ_T(\vx)\vJ_T(\vy)}{|\vx-\vy|}d\vx d\vy.
\end{equation}

\section{Instability with Breit Potential}\label{sec3}

\subsection{Static nuclei}

We now consider a system of $N$ (interacting) electrons in the
external electric field of $K$ static nuclei. There is no external
magnetic field but a self-generated one which is approximately
accounted for by the Breit potential. The energy is now
\begin{equation}\label{energy}
  \E_N(\psi,\vR)=\sprod{\psi}{(\sum_{\mu=1}^{N}D_{\mu}+\alpha (V_c-B))\psi}
\end{equation}
where
\begin{equation}\label{Breit}
  B= \sum_{\mu<\nu}^{N}\frac{1}{2|\vx_{\mu}-\vx_{\nu}|}
  \left(\sum_i\alpha_{i,\mu}\otimes\alpha_{i,\nu} +
  \frac{\valpha_{\mu}\cdot(\vx_{\mu}-\vx_{\nu})
  \otimes \valpha_{\nu}\cdot(\vx_{\mu}-\vx_{\nu})}{|\vx_{\mu}-\vx_{\nu}|^2}\right)
\end{equation}
and $V_c$ is the Coulomb potential defined in (\ref{Coulomb}).
$\vR$ denotes the $K$-tuple \((\vR_1,\ldots,\vR_K)\) of pairwise
different nuclear positions and $D_{\mu}=D_{\mu}(\vA\equiv 0)$. As
before $\psi$ belongs to \(\D_N\subset\H_N\). The interaction
$-\alpha B$ is usually derived from the corresponding interaction
in the Darwin Hamiltonian by the quantization
\(\vp/m\mapsto\valpha\) \cite{LandauLifshitz1971} or from QED:
treating the interactions of the electrons with the quantized
radiation field in second order perturbation theory leads to a
shift of the bound state energy levels approximately given by
\(-\alpha\sprod{\psi}{B\psi}\) \cite{BetheSalpeter1957}. Important
for our purpose is that
\begin{equation}\label{Breit-cc}
 \sprod{\psi}{B\psi} + \left(\begin{array}{c}\mbox{self-energy \&}\\
  \mbox{exchange terms}\end{array}\right)= \frac{1}{2}\int
  \frac{\vJ_T(\vx)\vJ_T(\vy)}{|\vx-\vy|} d\vx d\vy
\end{equation}
for any Slater determinant \(\psi=\psi_1\wedge\ldots\wedge\psi_N\)
of orthonormal functions $\psi_{\mu}$ (see the proof of
Theorem~\ref{theorem2}).

We are interested in the lowest possible energy
\begin{equation*}
   E_{N,K} = \inf \E_N(\psi,\vR)
\end{equation*}
where the infimum is taken over all \(\psi\in \D_N\) with $\|\psi\|=1$
and all $K$-tuples \((\vR_1,\ldots,\vR_K)\) with \(\vR_j\neq\vR_k\)
for $j\neq k$.  Our second main result is

\begin{theorem}\label{theorem2}
  There exists a constant $C$ such that for all \(\alpha>0,\ m\geq 0, K\in\N\) and
  \(Z_1,\ldots,Z_K\in\R_{+}\) \[E_{N,K}=-\infty\] whenever
  \(N,\sum Z_{\kappa}\geq C\max(\alpha^{-3/2},1)\). If \(\sum
  Z_{\kappa}^2\geq 1\) it suffices that
  \(C=5\cdot 10^4\) or - when $\alpha^{-1}=137$ - that \(N=\sum Z_\kappa\geq 3.4\cdot 10^7\).
\end{theorem}

\noindent
{\em Remarks.}
\begin{enumerate}
\item Similar as in Section~\ref{sec1}, $V_c$ and hence the condition
  on $\sum Z_{\kappa}$ may be dropped. Then there is instability for
  $N\geq C\max(\alpha^{-3/2},1)$. It is for completeness of the model
  we keep $V_c$ in this section.
\item Without $B$ the energy is proven to be non-negative \(\alpha
  Z_{\kappa}\leq2/\pi\) for all $\kappa$ and if \(\alpha\leq 1/94\)
  \cite{LiebYau1988} (see also \cite{Liebetal1997}). One expects
  however stability even for \(\alpha Z_{\kappa} \leq
  2\left(\frac{2}{\pi}+\frac{\pi}{2}\right)^{-1}\) $\alpha\leq 0.12$
  \cite{Evansetal1996, BalinskyEvans1998b}, which would cover the
  atomic numbers of all known elements.
\end{enumerate}

At least partly this theorem can be understood from Corollary
\ref{corollary}, Equation (\ref{minimum}) and Equation
(\ref{Breit-cc}). \medskip

\begin{proof}[Proof of Theorem~\ref{theorem2}]
  To begin with we prove (\ref{Breit-cc}). Let \(\psi =
  \psi_1\wedge\ldots\wedge\psi_N\) with
  \(\sprod{\psi_{\mu}}{\psi_{\nu}}= \delta_{\mu\nu}\) and let
  \(\vJ(\vx)= \sum_{\mu=1}^{N}
  \psi_{\mu}^{+}(x)\valpha\psi_{\mu}(x)\) be the current
  density of $\psi$. Note that \(\widehat{J}_{T,i}(\vp)=
  \sum_{j=1}^{3}(\delta_{ij}- \frac{p_ip_j}{p^2})\widehat{J}_{j}(\vp)\) and that
\begin{equation*}
  \mathfrak{F}\frac{4\pi}{p^2}(\delta_{ij}-\frac{p_ip_j}{p^2})
   = \frac{1}{2|x|}\left(\delta_{ij}+\frac{x_ix_j}{x^2}\right).
\end{equation*}
With $B(\vx)$ defined by
\begin{equation*}
  B(x)= \frac{1}{2|x|}\sum_{i,j}\alpha_i
  \left(\delta_{ij}+\frac{x_ix_j}{x^2}\right)\alpha_j =
  \frac{1}{2|x|}\left(\sum_i\alpha_i\otimes\alpha_i+
  \frac{\valpha\cdot\vx\otimes
  \valpha\cdot\vx}{|\vx|^2}\right)
\end{equation*}
it follows that
\begin{equation}\begin{split}\label{Breit=cc}
  \frac{1}{2}\int
  \frac{\vJ_T(\vx)\vJ_T(\vy)}{|\vx-\vy|}
  d\vx d\vy
  &= \frac{1}{2}
  \sum_{\mu,\nu}\sprod{\psi_{\mu}\otimes\psi_{\nu}}{B(x-y)\psi_{\mu}\otimes\psi_{\nu}}\\
  &= \sprod{\psi}{B\psi}+ \frac{1}{2}
  \sum_{\mu,\nu}\sprod{\psi_{\mu}\otimes\psi_{\nu}}{B(x-y)\psi_{\nu}\otimes\psi_{\mu}}.
\end{split}
\end{equation}
which is equation (\ref{Breit-cc}). Similar as in the proof of
Theorem~\ref{theorem1} it suffices to consider the case $m=0$ and to
find a Slater determinant \(\psi=\psi_1\wedge\ldots\wedge\psi_N\) and
nuclear positions  \(\vR_1,\ldots,\vR_K\) such that
\(\E_N(\psi,\vR)<0\).

{\em Choice of the nuclear positions.} A beautiful argument given in
\cite{Liebetal1997} show that, after moving some electrons or nuclei
far away from all others
\begin{equation*}
  \sprod{\psi}{V_c\psi}\leq \eps+\frac{1}{2N^2}\sum_{\mu,\nu} \int
  \frac{|\psi_{\mu}(\vx)|^2 |\psi_{\nu}(\vy)|^2}{|\vx-\vy|}
  d\vx d\vy
\end{equation*}
for suitably chosen nuclear positions. Here $\eps>0$ is the (arbitrary
small) contribution of the particles moved away. The second term can
be dropped if \(\sum_{\kappa=1}^{K}Z_{\kappa}^2\geq 1\). We use the
inequality obtained in \cite{Tix1997} to estimate it from above and find
\begin{equation}\label{bpf1}
  \sprod{\psi}{V_c\psi}\leq \eps + \const\frac{1}{N}\sum_{\mu=1}^{N}
  \sprod{\psi_{\mu}}{D\psi_{\mu}}.
\end{equation}
The number $N$ of remaining electrons obeys $N<\sum Z_{\kappa}+1$
which is the reason for the assumption on \(\sum Z_{\kappa}\). Of
course the choice of the nuclear positions depends on $\psi$, which
has not been specified yet.

Define one-particle orbitals $\psi_{\mu}$ and currents $\vJ_{\mu}$ and
$\vJ_{0}$ exactly as in the proof of Theorem~\ref{theorem1} with
$\ve$ being an arbitrary unit vector in $\R^3$. The convergence
\(\widehat{\vJ}_{\mu}(\vp)\rightarrow\widehat{\vJ}_{0}(\vp)\) as
$\lambda\rightarrow\infty$ now implies that
\begin{equation}\label{bpf2}\begin{split}
\frac{1}{2}\int \frac{\vJ_T(\vx)\vJ_T(\vy)}{|\vx-\vy|}
d\vx d\vy &= N^2\left[\frac{1}{2} \int \frac{\vJ_{0,T}(\vx)\vJ_{0,T}(\vy)}{|\vx-\vy|}
d\vx d\vy + O(\lambda^{-1})\right] \\ &\geq c_2 N^2
\end{split}
\end{equation}
for \(\lambda\geq\lambda_0\), where $\lambda_0$ and $c_2>0$ are independent of $N$.

To estimate the sum of exchange- and self-energy terms in
(\ref{Breit=cc}) notice that
\begin{equation}\label{bpf3}
  \sprod{\psi_{\mu}\otimes\psi_{\nu}}{B(x-y)\psi_{\nu}\otimes\psi_{\mu}} =
  \int\frac{4\pi}{p^2}|\widehat{\vJ}_{\mu\nu,T}(\vp)|^2 d\vp,
\end{equation}
where \(\vJ_{\mu\nu}(\vx)= \psi_{\mu}^{\ast}(x)\valpha\psi_{\nu}(x)\).
After writing \(\widehat{\vJ}_{\mu\nu}(\vp)\) as an integral
in Fourier space in terms of $u_{\mu}$ and $u_{\nu}$ similar as in
(\ref{apf5}) it is easily seen, using the support properties of $u_{\mu}$ and
$u_{\nu}$, that
\begin{equation}\label{bpf4}
  |\widehat{\vJ}_{\mu\nu,T}(\vp)|^2\leq|\widehat{\vJ}_{\mu\nu}(\vp)|^2\leq
   3(2\pi)^{-3}\chi(|\vp+\vn_{\mu}-\vn_{\nu}|
   \leq\sqrt{3}).
\end{equation}
The $N$ balls \(B(\vn_{\nu},\sqrt{3}),\ \nu=1,\ldots,N\) all lie in the ball
$B(0,N^{1/3})$ and cover a given point at most, say, $4^3=64$ times
(replace the balls by cubes with side $2\sqrt{3}$). Therefore (\ref{bpf4})
implies
\begin{equation*}
  \sum_{\nu=1}^{N} |\widehat{\vJ}_{\mu\nu}(\vp)|^2\leq
  192(2\pi)^{-3}\chi(|\vp+\vn_{\mu}|<N^{1/3})
  \leq \frac{24}{\pi^3}\chi(|\vp|<2N^{1/3})
\end{equation*}
which in conjunction with (\ref{bpf3}) gives
\begin{equation}\label{bpf5}
  \frac{1}{2}\sum_{\mu,\nu}\sprod{\psi_{\mu}\otimes\psi_{\nu}}{B(x-y)\psi_{\nu}\otimes\psi_{\mu}}
  \leq\frac{384}{\pi}\ N^{4/3}.
\end{equation}

Rewriting the energy using (\ref{Breit=cc}) and inserting the
estimates (\ref{bpf1}), (\ref{apf8}), (\ref{bpf2}) and (\ref{bpf5}) we arrive at
\begin{equation*}
  \E_N(\psi,\vR)\leq c_1 (1+\alpha)N^{4/3}-c_2\alpha N^{2},\hspace{3em}c_2>0
\end{equation*}
which is negative for \(N>\const\ \max(\alpha^{-3/2},1)\). This proves the theorem.
\end{proof}

\noindent
For small $N$ and small $\alpha$ there is stability. A similar result
for the energy in Section 1 was proved in \cite{Liebetal1997}.

\begin{theorem}\label{theorem3}
Suppose \(\tilde{\alpha}\leq 1/94\), \(\max_\kappa\ Z_{\kappa}\leq
2/\pi\,\tilde{\alpha}^{-1}\) and \(N-1\leq
2(2/\pi+\pi/2)(\alpha^{-1}-\tilde{\alpha}^{-1})\). Then
\(E_{N,K}\geq 0\). Inserting \(\tilde{\alpha}= 1/94\) and
\(\alpha=1/137\) we find stability for $N\leq 39$ and \(\max\
Z_{\kappa}\leq 59\).
\end{theorem}

\begin{proof} Since \(B(x)\leq 2/|x|\) on \(\C^4\otimes\C^4\) and
  \(1/|x|\leq \delta^{-1}D\) on \(\Lambda_{+}L^2(\R^3;\C^4)\)
  where \(\delta=2(2/\pi+\pi/2)\) \cite{Tix1997} one has by the
  symmetry property of the states in $\H_N$
\begin{equation}\label{spf1}
  B\leq \frac{N-1}{\delta}\sum_{\mu=1}^{N}D_{\mu}\hspace{3em}\hbox{on}\ \H_N.
\end{equation}
Furthermore
\begin{equation}\label{spf2}
  V_c\geq -\frac{1}{\tilde{\alpha}}\sum_{\mu=1}^{N}D_{\mu}\hspace{3em}\hbox{on}\ \H_N
\end{equation}
for all $\tilde{\alpha}>0$ with \(\tilde{\alpha}\max\ Z_{\kappa}\leq
\frac{2}{\pi}\) and \(\tilde{\alpha}q\leq 1/47\) by \cite{LiebYau1988},
where the number $q$ of spin states may be set equal 2
\cite{Liebetal1997}. Inserting (\ref{spf1}) and (\ref{spf2}) in the
energy proves the theorem.
\end{proof}

\subsection{Dynamic nuclei}
\label{subsec3}

Making the nuclei dynamical would improve stability if their
kinetic energy were the only term we added to (\ref{energy}).
However if the nuclei are relativistic spin 1/2 particles like the
electrons and if the Breit-potential couples all pairs of
particles, taking their charges into account, then the instability
will actually become worse.

Let us illustrate this for a system of $N$ electrons and $K$ identical
nuclei of spin 1/2 and atomic number $Z>0$. These nuclei are described
by vectors in the positive energy subspace of the free Dirac operator
with the mass $M>0$ of the nuclei. To prove instability we
adopt the strategy of the proof of Theorem~\ref{theorem2} and thus
assume $M=0$ and $m=0$. As a trial-wave function we take
\begin{equation*}
\psi = (\psi_1\wedge\ldots\wedge\psi_{N})\otimes(\phi_1\wedge\ldots\wedge\phi_{K})
\end{equation*}
where $\psi_{\mu}$ is defined by equations (\ref{apf1.5}) and
(\ref{apf2}) and $\phi_{\kappa}$ is defined like $\psi_{\kappa}$
except that $\ve$ and $N$ are replaced by $-\ve$ and $K$ respectively.
It follows that in the limit \(\lambda\rightarrow\infty\) we get $N+K$
(charge-) currents, the nuclear ones being larger than the electronic
ones by a factor of $Z$ but otherwise identical. The Breit
interactions thus gives a negative contribution to the energy of order
\(\alpha(N+ZK)^2\). While the parallel currents of the $N+K$ particles
add up, the opposite charges of the electrons and nuclei cancel
themselves. In fact for $\psi$ defined as above
\begin{equation}\begin{split}
\sprod{\psi}{V_c\psi} &\leq  \sum_{\mu<\nu}^N\int d\vx
d\vy\frac{|\psi_{\mu}(\vx)|^2|\psi_{\nu}(\vy)|^2}{|\vx-\vy|} \\
& \quad + Z^2\sum_{\kappa<\sigma}^K\int d\vR_1
d\vR_2\frac{|\phi_{\kappa}(\vR_1)|^2|\phi_{\sigma}(\vR_2)|^2}{|\vR_1-\vR_2|}
\\
& \quad+ Z\sum_{\kappa=1}^K\sum_{\mu=1}^N\int d\vx
d\vR\frac{|\psi_{\mu}(\vx)|^2|\phi_{\kappa}(\vR)|^2}{|\vx-\vR|}\\
&=
\left[\frac{N(N-1)}{2}+Z^2\frac{K(K-1)}{2}-NKZ\right](I+O(\lambda^{-1}))\\
&= \left[(KZ-N)^{2}-KZ^2-N\right](I/2+O(\lambda^{-1})),
\end{split}
\end{equation}
where $I$ is the limit of the above double integrals as
$\lambda\rightarrow \infty$. Hence \(\sprod{\psi}{V_c\psi}\) is
negative, e.g., if $KZ=N$ and \(\lambda\) is large. To achieve
this in the static case we had to choose the nuclear positions
properly. It is instructive to recall how this was done. The total
energy is bounded from above by
\(c_1(N^{4/3}+K^{4/3})-c_2\alpha(N+KZ)^2,\ c_2>0,\) for $N=KZ$ and
$\lambda$ large, and is therefore negative for $N=KZ$ large
enough.

\section{Stability and Instability with Quantized Radiation Field}\label{sec4}

Instability for the model with classical magnetic field implies
instability for the model with quantized radiation field without
UV-cutoff. In fact, for each classical magnetic field there is a
coherent state of photons which reproduces the classical field as
far as the energy is concerned. If an UV cutoff is introduced the
relativistic scale invariance of the energy is broken and
stability of the first kind is restored. The lower bound depends
on the cutoff and goes to $-\infty$ as the cutoff is removed.

The state of the system is now described by a vector \(\Psi\in
\H_N\otimes\F\) where $\F$ denotes the bosonic Fock-space over
\(L^2(\R^3)\otimes\C^2\), the factor $\C^2$ accounting for the two
possible polarizations of the transversal photons, and the total
energy of $\Psi$ is
\begin{align*}
  \E_N^{\text{qed}}(\Psi) &=
  \sprod{\Psi}{\sum_{\mu=1}^{N}[\valpha_{\mu}\cdot(-i\nabla_{\mu} +
  \alpha^{1/2}\vA(\vx_{\mu}))+\beta_{\mu}m]\Psi} \\
  &\quad+\sprod{\Psi}{(1\otimes H_f)\Psi}\\
  H_f &= \sum_{\lambda=1}^{2}\int d\vk |\vk|
  a_{\lambda}^{\dagger}(\vk) a_{\lambda}(\vk),
\end{align*}
where
\begin{align*}
  \vA(\vx) &:= \sum_{\lambda=1}^{2}\int dk
  \left[\ve_{\lambda}(\vk)e^{i\vektor{kx}}\otimes
  a_{\lambda}(\vk) + \ve_{\lambda}(\vk)e^{-i\vektor{kx}}\otimes
  a_{\lambda}^{\dagger}(\vk)\right]\\
  &=: \vA^{+}(\vx)+\vA^{+}(\vx)^{*}
\end{align*}
is the quantized vector potential in Coulomb gauge. The operators $a_{\lambda}(\vk)$
and $a_{\lambda}^{\dagger}(\vk)$ are creation and annihilation
operators acting on $\F$ and obeying the CCR
\begin{equation*}
  [a_{\lambda}(\vk_1),a_{\mu}^{\dagger}(\vk_2)] =
  \delta_{\mu\nu}\delta(\vk_1-\vk_2),\hspace{3em}
  [a_{\lambda}^{\sharp}(\vk_1),a_{\mu}^{\sharp}(\vk_2)] = 0
\end{equation*}
where $a_{\lambda}^{\sharp}=a_{\lambda}$ or
$a_{\lambda}^{\dagger}$, and the two polarization vectors
$\ve_{\lambda}(\vk)$ are orthonormal and perpendicular to $\vk$
for each $\vk\in \R^3$. We use $dk$ as a short hand for
\((2\pi)^{-3/2}(2|\vk|)^{-1/2}d\vk\), and the subindex of
\(\valpha_{\mu},\ \nabla_{\mu}\) and $\beta_{\mu}$ indicates that
these one particle operators act on the $\mu$-th particle.  While
we used Gaussian units in Section~\ref{sec2} and \ref{sec3} we now
work with Heaviside Lorenz units.

\begin{lemma}\label{lemma}
  For each \(\vA_{cl}\in\mathcal{A}\cap L^2(\R^3)\) there exists a
  vector \(\theta\in \F\) (coherent state) such that
  \[ \E_N^{\text{qed}}(\psi\otimes\theta) = \E_N(\psi,\vA_{cl}) \]
  for all \(\psi\in \D_N.\)
\end{lemma}

\begin{proof}
Pick \(\vA_{cl}\in \mathcal{A}\cap L^2(\R^3)\) and define
\(\eta_{\lambda(\vk)}=(|\vk|/2)^{1/2} \ve_{\lambda}(\vk)\cdot\widehat{\vA}_{cl}(\vk)\)
so that \(\vA_{cl}(\vx)=\vA_{cl}^{+}(\vx)+\vA_{cl}^{+}(\vx)^{\ast}\) with
\begin{equation}\label{qapf1}
  \vA_{cl}^{+}(\vx)=\sum_{\lambda=1}^{2}\int dk \eta_{\lambda}(\vk)\ve_{\lambda}(\vk)e^{i\vk\vx}.
\end{equation}
Next set
\begin{equation*}
  \Pi(\eta) := i\sum_{\lambda=1}^{2}\int d\vk
  \left[\overline{\eta_{\lambda}(\vk)}a_{\lambda}(\vk)+
  \eta_{\lambda}(\vk)a_{\lambda}^{\dagger}(\vk)\right]
\end{equation*}
and \(\Theta = e^{-i\Pi(\eta)}\Omega\in\F\). $\Theta$ is called a
coherent state, it is normalized and most importantly it is an
eigenvector of all annihilation operators
\begin{equation}\label{qipf2}
  a_{\lambda}(\vk)\Theta = \eta_{\lambda}(\vk)\Theta.
\end{equation}
>From (\ref{qapf1}), (\ref{qipf2}) and the
definition of $\eta_{\lambda}(\vk)$ it follows that
\begin{equation*}
  \valpha_{\mu}\vA^{+}(\vx_{\mu})\psi\otimes\Theta =
  \left(\valpha_{\mu}\vA^{+}_{cl}(\vx_{\mu})\otimes\vektor{1}\right) \psi\otimes\Theta
\end{equation*}
and
\begin{equation*}
\sprod{\Theta}{H_f\Theta} = \int d\vk |\vk|
\sum_{\lambda}|\eta_{\lambda}(\vk)|^2=\frac{1}{2}\int d\vk k^2|\widehat{\vA}_{cl}(\vk)|^2.
\end{equation*}
Inserting this in the energy proves the theorem.
\end{proof}

If an ultraviolet cutoff is introduced in the field operator
$\vA(\vx)$ then stability of the first kind is restored for all
$N$ and a certain range of values for $\alpha$ and $Z_{\kappa}$.
This follows from \cite[Lemma I.5]{Bachetal1998a} and
\cite[Theorem 1]{Liebetal1997}.

\section*{Acknowledgement}
\label{sec:ack}

It is a pleasure to thank Heinz Siedentop for many discussions,
and Arne Jensen, Jan Philip Solovej and Erik Skibsted for the
hospitality at Aarhus University in August 97, where this work was
begun. M.~G.~ also thanks Michael Loss for clarifying discussions.
This work was partially supported by the European Union under
grant ERB4001GT950214 and under the TMR-network grant FMRX-CT
96-0001.

\appendix
\renewcommand{\theequation}{\Alph{section}.\arabic{equation}}

\section*{Appendix}
\addcontentsline{toc}{section}{Appendix}

\setcounter{section}{1} \setcounter{equation}{0}
\setcounter{theorem}{0}

To obtain the numerical values for the constants in
Corollary~\ref{corollary} and Theorem~\ref{theorem2} we follow the
proof of Theorem~\ref{theorem2}, up to a few modifications and
explicitly evaluate the constants in this proof.

The main modifications are that the two-spinor $u$ is now defined in terms
of the (normalized) characteristic function of the ball with radius
$1/2$ contained in the unit cube $\{\vp|0\leq p_i\leq 1\}$ and that
the 4-spinors $\psi_{2\mu-1}$ are defined in terms of the
$\psi_{2\mu}$'s by interchanging the components of $\vektor{u}$, while
$\vn_{2\mu-1}$ runs over the $N/2$ or - if $N$ is odd - the
$(N+1)/2$ lattice sites of $\Z^3$ closest to the origin. The balls
simplify the computation of \(\widehat{\vJ}_0(\vp)\) and the double
occupation \(\vn_{2\mu-1}=\vn_{2\mu}\) reduces the
kinetic energy. To begin with we note that the $n$ unit cubes of
the lattice $\Z^3$ which are closest to the origin, all fit in a ball of
radius
\[ n^{1/3}\left(\frac{3}{4\pi}\right)^{1/3} + \sqrt{3}. \]
In particular the $N/2$ or $(N+1)/2$ unit cubes containing the
supports of the spinors \(\psi_{\mu},\ \mu=1,\ldots,N\) all lie in the
ball of radius $bN^{1/3}$ centered at $\lambda N^{1/3}\ve$ where
$b=1/2$ if $N\geq 1.2\cdot 10^7$, $b=3/5$ if $N\geq 5\cdot 10^3$ and
$b=\sqrt{3}$ if $N\geq 1$ (the ball of radius $\sqrt{3}n^{1/3}$
contains never less than $n$ lattice cubes).  This replaces equation
(\ref{apf2}) and implies, together with equations (\ref{apf5}) and
(\ref{apf6}), that
  \[ |\widehat{\vJ}_{\mu}(\vp) - \widehat{\vJ}_{0}(\vp)| \leq
  \frac{6b}{\lambda-b}|\widehat{\vJ}_{0}(\vp)|,\hspace{3em} \lambda>b.\]
Using this and \(|\widehat{\vJ}_{0}(\vp)|=1/2(2\pi)^{-3/2}(1-p)^2(2+p)\)
one finds
\begin{equation}\label{num1}
\begin{split}
  \int\frac{\vJ_T(\vx)\vJ_T(\vy)}{|\vx-\vy|}d\vx d\vy &=
  \sum_{\mu,\nu=1}^N\int d\vp\frac{4\pi}{\vp^2}
  \widehat{\vJ}_{\mu}^{\ast}(\vp)T\widehat{\vJ}_{\nu}(\vp)\\
  &\geq N^2\left[\int d\vp\frac{4\pi}{p^2}
  \widehat{\vJ}_{0}^{\ast}(\vp)T\widehat{\vJ}_{0}(\vp) -
  \frac{12b}{\lambda-b}\int d\vp\frac{4\pi}{p^2}
  |\widehat{\vJ}_{0}(\vp)|^2\right]\\
  &= N^2 \left[1-\frac{18b}{\lambda-b}\right]\frac{11}{35\pi},
\end{split}
\end{equation}
where $T$ is the $3\times 3$ matrix with the components \(T_{ij}=\delta_{ij}-p_ip_j/p^2\).
This replaces (\ref{bpf2}). We proceed to bound the self-energy and
exchange terms. Inequality (\ref{bpf4}) becomes
\begin{equation}\label{num2}
  |\widehat{\vJ}_{\mu\nu}(\vp)|^2\leq
  3(2\pi)^{-3}\chi(|\vp+\vn_{\mu}-\vn_{\nu}| \leq 1),
\end{equation}
because the support of $u$ now has diameter $1$ not $\sqrt{3}$. Since
the $N$ balls \(B(\vn_{\nu},1)\) cover a given point at most 8
times (recall that now \(\vn_{2\nu-1}=\vn_{2\nu}\))
inequality (\ref{num2}) leads to the bound
\begin{equation}\label{num3}
  \frac{1}{2}\sum_{\mu,\nu} \sprod{\psi_{\mu}\otimes\psi_{\nu}}{B(x-y)\psi_{\nu}\otimes\psi_{\mu}}
  \leq \frac{48}{\pi}bN^{4/3}
\end{equation}
improving (\ref{bpf5}). The kinetic energy is bounded by
\((\lambda+b)N^{4/3}\) and \(\sprod{\psi}{V_c\psi}\leq 0\) since
\(\sum Z_{\kappa}=N\) and \(\sum Z_{\kappa}^2\geq 1\). In conjunction
with (\ref{num1}) and (\ref{num3}) this gives
\begin{equation}\label{num4}
\E_N(\psi) \leq N^{4/3}\left[\lambda+b +\frac{48}{\pi}b\alpha
  -\alpha N^{2/3}\frac{11}{70\pi}\left(1-\frac{18b}{\lambda-b}\right) \right].
\end{equation}
For \(b=1/2,\ \alpha^{-1}=137\) and the optimal $\lambda$ this is
negative for \(N\geq 3.4\cdot10^7\). For \(b=3/5\) and $\alpha>0$
arbitrary this is negative for \(N\geq C\max(\alpha^{-3/2},1)\) with
$C=43859$ where $\lambda$ was chosen to minimize $C$. This explains
the numbers in Theorem~\ref{theorem2}.

Now drop the term \((48/\pi)b\alpha\) in equation (\ref{num4})
which was due to the exchange- and self-energy terms. By equation
(\ref{minimum}) what we are left with is a upper bound for
\(\inf_{\psi,
  \vA\in\mathcal{A}} \E_N(\psi,\vA)\). It
is negative for $b=\sqrt{3}$, the optimal $\lambda$ and
\(\alpha^{3/2}N\geq 134'863\), or for $b=1/2$ the optimal
$\lambda$, $\alpha^{-1}=137$, and \(N\geq 3.4\cdot 10^7\). This
explains the numbers in the remark after
Corollary~\ref{corollary}.


\end{document}